\documentclass[]{template}

\usepackage{algorithm}
\usepackage{algpseudocode}
\usepackage{gensymb}
\usepackage{lipsum}
\title{JAX-BEM: Gradient-Based Acoustic Shape Optimisation via a Differentiable Boundary Element Method}

\author[1,2]{James Hipperson}
\author[1]{Jonathan A. Hargreaves}
\author[1]{Trevor J. Cox}
\correspondingauthor{j.r.hipperson@edu.salford.ac.uk}{Hipperson et al.}

\affil[1]{Acoustics Research Centre, University of Salford, UK}
\affil[2]{Funktion-One Research Ltd. Dorking, UK}

\begin{document}
\maketitle

\begin{abstract}
Engineering structures are increasingly designed using numerical optimisation. However, traditional optimisation methods can be challenging with multiple objectives and many parameters. In machine learning, stable training of artificial neural networks with millions or billions of parameters is achieved using automatic differentiation frameworks such as \texttt{JAX} and \texttt{Pytorch}. Because these frameworks provide accelerated numerical linear algebra with automatic gradient tracking, they also enable differentiable implementations of numerical methods to be built. This facilitates faster gradient-based optimisation of geometry and materials, as well as solution of inverse problems. We demonstrate \texttt{JAX-BEM}, a differentiable Boundary Element Method (BEM) solver, showing that it matches the error of existing BEM codes for a benchmark problem and enables gradient-based geometry optimisation. Although the demonstrated examples are for acoustic simulations, the concept could be readily extended to electromagnetic waves. 

\end{abstract}

\section{Introduction}\label{sec:introduction}

In acoustics, there is a need to characterise radiation from devices such as loudspeakers, scatterers or metamaterials, often in free-field conditions equivalent to an unbounded domain. Domain-based numerical methods, such as the Finite Element Method (FEM) and Finite Difference Time Domain (FDTD), can simulate this using Perfectly Matched Layers (PMLs), infinite elements, or Absorbing Boundary Conditions (ABCs) \cite{Sakuma2014}, but the air domain must still be meshed, which creates many degrees of freedom as the number of elements scales with domain size (approximately quadratically in 2D with area and cubically in 3D with volume). The Boundary Element Method (BEM), on the other hand has significant advantages in this use case because it only requires meshing of boundaries.

Using automatic differentiation frameworks to build differentiable numerical methods is a recent area of development, although adjoint methods are related \cite{cea_conception_1986}. Differentiability enables gradient-based optimisation, inverse problems such as material parameter characterisation, and tight integration with machine learning methods. A differentiable FEM implementation was demonstrated by Xue et al. in 2023 \cite{xue_jax-fem_2023} and used for acoustic optimisation by Borrel-Jensen and Bjorgaard in 2025 \cite{borrel-jensen_differentiation_2025}. In this paper, we demonstrate gradient-based geometry optimisation using JAX-BEM for a loudspeaker horn in an unbounded domain. 

\section{Background}

\subsection{Boundary Element Method (BEM)}

BEM utilises the Huygens-Fresnel principle and the Kirchhoff-Helmholtz boundary integral to efficiently compute solutions to problems within a domain using only data defined on its boundaries. In unbounded domains, the Sommerfeld radiation condition can be used to show that the contribution from an imagined very distant outer boundary is zero, hence only data on the boundary of the obstacle is required, leading to a very efficient numerical representation. 

BEM is formulated in two phases: (1) A system of equations is set up relating the complex amplitudes of a distribution of sources on the boundary to pressure (or its gradient) elsewhere on the boundary $S$. The boundary conditions are then inserted and this is solved numerically assuming a mesh of elements and interpolation functions. (2) The solution from step 1 (being pressure and/or velocity on the boundary) is used to calculate the scattered pressure at points in the domain $\Omega$.

Both stages are based on via the Kirchhoff-Helmholtz boundary integral equation:

\begin{equation}
\label{eq:kirchoff-helmholtz}
  p_\mathrm{s}(\mathbf{r}) = \oint_S \left[
    G(\mathbf{r}, \mathbf{r}') \frac{\partial p_\mathrm{t}(\mathbf{r}')}{\partial n'}
    - p_\mathrm{t}(\mathbf{r}') \frac{\partial G(\mathbf{r}, \mathbf{r}')}{\partial n'}
  \right] dS',
\end{equation}

Here $p_\mathrm{s}$ is scattered pressure and $p_\mathrm{t} = p_i + p_\mathrm{s}$ is total pressure, where $p_\mathrm{i}$ is some incident pressure field that is required if a scattering problem is being modelled. $\mathbf{r}\in\Omega$ and $\mathbf{r}'\in S$ are domain and boundary points and $G$ is the free space Green's function at wavenumber $k$:

\begin{equation}
\label{eq:greens_function}
    G(\mathbf{r}, \mathbf{r}') = \frac{\mathrm{e}^{\mathrm{i}k|\mathbf{r} - \mathbf{r}'|}}{4\pi |\mathbf{r} - \mathbf{r}'|}
\end{equation}

Slightly different formulations arise for internal and free-field (exterior) domains and varying boundary conditions. In either case, collocating at $N$ boundary nodes yields a dense system of matrix equations size $N^2$ to be solved for the unknown boundary quantity. Other variations in formulation exist such as Burton-Miller \cite{burton_application_1971} and CHIEF \cite{schenck_improved_1968} to address non-uniqueness. 

\subsection{Automatic Differentiation (autodiff)}

The methods that made deep learning possible are backpropagation \cite{rumelhart_learning_1986} and automatic differentiation \cite{griewank_autodiff_2008}. These are used to train models with millions or billions of parameters \cite{brown_language_2020}, which was not feasible with previous optimisation methods. Automatic differentiation works by storing a computational graph of an algorithm step-by-step and simultaneously constructing a second computational graph of derivatives at runtime. This enables gradients to be tracked through an algorithm automatically. Backpropagation (more generally, reverse-mode automatic differentiation) is the process by which gradients are propagated from the end of an algorithm, typically a scalar loss function, backwards through the algorithm with respect to the input parameters using the chain rule. Software frameworks facilitate this, e.g.  \texttt{JAX}~\cite{jax2018github} and \texttt{Pytorch}~\cite{Pytorch}. Beyond neural network weights and biases, automatic differentiation can also be used to build differentiable implementations of numerical methods.

\section{Methodology}\label{sec:methodology}

Using the open source BEM project \texttt{bempp} \cite{betcke_bempp-cl_2021} as a reference, the operators were refactored to just-in-time (JIT) compiled vectorised \texttt{JAX} functions. Because \texttt{JAX} can automatically track gradients, the only manual handling of differentiability was to compute the adjacency information of a mesh outside the geometry optimisation loop (as this necessarily contained non-differentiable control flow statements), and to use implicit function theorem (implicit differentiation) to avoid "unrolling" the iterative solver (\texttt{GMRES}) \cite{saad_gmres_1986}, which would be inefficient and use a lot of memory. 

\texttt{JAX} supports accelerated linear algebra operations on CPU, GPU and TPU hardware via the \texttt{OpenXLA} backend. This is useful as the boundary-to-domain propagation phase of BEM is "embarrassingly parallel" and highly amenable to GPU and TPU computation. 

\subsection{Non-differentiable statements}

While \texttt{JAX} closely follows the functionality of \texttt{Numpy} \cite{harris2020array}, it is recommended to follow a pure function approach, avoid loops and control flow statements. This ensures the \texttt{JAX} program is JIT (just-in-time) compilable, and differentiable. Loops are problematic as they must be unrolled to track gradients, and loops with an indeterminate number of iterations are not differentiable. Branching statements such as \texttt{if} or \texttt{switch} are not differentiable and cannot be used within the optimisation loop. Keeping these factors in mind while programming, the automatic handling of gradient tracking in \texttt{JAX} makes writing differentiable programs relatively straightforward. 

A specific difficulty with BEM is computation of the adjacency matrix and handling of singular integration routines. Special handling is required for element self-interaction (singular) and adjacent element interaction (nearly-singular). Here, branching statements cannot be avoided, so adjacency and selection of integration types is re-computed once per iteration before the differentiable sequence (Alg. \ref{alg:overview}).

\begin{algorithm}
\caption{Geometry optimisation for scattering}\label{alg:overview}
\begin{algorithmic}
\Require $p_{i}, mesh_\mathrm{init}, k$
\Ensure $mesh$

\State $Vertices, Elements \gets$ load$(mesh_\mathrm{init})$

\For{$iteration = 1$ to K}
    \If {$iteration>0$}
        \State $Vertices, Elements \gets $load$(mesh)$
    \EndIf
    \State $p_{s}, A \gets$ forward$(k, p_{i}, Vertices, Elements)$
    \State $g \gets$ loss$(p_{s})$
    \State $\nabla_V \gets$ backward$(p_\mathrm{s},A,g)$
    \State $vertices \gets vertices - H^{-1} \cdot \nabla_V$
    \State $mesh \gets vertices$

\EndFor
\State \Return $mesh$
\end{algorithmic}
\end{algorithm}

Here $A$ is the boundary integral operator matrix (left hand side), $mesh_\mathrm{init}$ is the initial starting mesh and $H^{-1}$ is the inverse Hessian approximated by the L-BGFS optimiser \cite{liu_limited_1989}. Forward is shown in Alg. \ref{alg:fwd} and backward in Alg. \ref{alg:bwd}. 

\subsection{Implicit differentiation}

\texttt{GMRES} is an iterative solver and this is problematic in differentiable computing for two reasons, it is: (1) iterative and runs in a loop, and (2) typically configured to stop at a target tolerance, which means that the number of iterations is indeterminate. \texttt{JAX} contains several useful tools to overcome these problems. Using \texttt{JAX.custom\_vjp} (Vector-Jacobian Product) we define the forward (primal) function and the backward function (pullback) to set up \emph{implicit} differentiation so that autodiff bypasses both instances of \texttt{GMRES}. \texttt{nondiff\_argnames} excludes non-differentiable arguments from gradient computation by treating them as constants. The forward function is:

\begin{algorithm}
\caption{BEM Forward function}\label{alg:fwd}
\begin{algorithmic}
\Require $k, p_\mathrm{i}, Vertices, Elements$
\Ensure $p_\mathrm{s}, A$
    \State $A \gets$ assemble$(k, Vertices, Elements)$
    \State $b \gets $ project$(p_\mathrm{i})$
    \State solve $Ax=b$ using GMRES
    \State $p_\mathrm{s} \gets x$
    \State \Return $p_\mathrm{s}, A$
\end{algorithmic}
\end{algorithm}

The backward function (Alg. \ref{alg:bwd}) is where implicit differentiation happens, yielding gradients with respect to $A$ and $b$, and by the chain rule $V$ (vertices). It does this by solving the adjoint system $A^H \lambda = g$. (Note, this is essentially an automated version of the adjoint state method \cite{cea_conception_1986}.)

\begin{algorithm}
\caption{Implicit \texttt{GMRES} Backward function}\label{alg:bwd}
\begin{algorithmic}
\Require $p_\mathrm{s}$, $A$, loss $g$
\Ensure Gradient $\nabla_V$
    \State solve $A^H \lambda = g$ using GMRES
    \State $\nabla_A \gets -\lambda p_\mathrm{s}^H$
    \State $\nabla_b \gets \lambda$
    \State $\nabla_V \leftarrow \left({\partial A / \partial V}\right)^{\top} \nabla_A$
    \State \Return $\nabla_V$
\end{algorithmic}
\end{algorithm}

The key is that by using implicit differentiation, autodiff never touches the iterative solver; it remains a black box. Only the inputs and outputs pass gradients. Any solver could be used in principle, but the \texttt{JAX} implementation of \texttt{GMRES} was used (\texttt{jax.scipy.sparse.linalg.gmres}) to enable JIT compilation, as both forward and backward functions are run many times in an optimisation loop. 

\subsection{Geometry optimisation}

The parameters to be optimised are the control points of splines that define the axial horn profile. The initial mesh is then deformed to follow the splines. This work extends prior approaches \cite{udawalpola_optimization_2011}, \cite{godinho_3d_2015}, \cite{schmidt_large-scale_2016}, \cite{andersen_3d_2022}, \cite{dong_efficient_2023} by using gradients in the optimisation. The L-BGFS optimiser was used, specifically \texttt{scipy.opimize.minimize}. 

\subsection{Loss (cost) function}

The loss function is a mean squared error loss against dB targets relative to the normalised on-axis response.This was set to $T_{\mathrm{in}} = -3\mathrm{dB}$ within the desired coverage region ($\pm 35 \degree$ Horizontal, $\pm 25 \degree$ Vertical) and $T_{\mathrm{out}} = -10\mathrm{dB}$ outside of the desired coverage region. 

\begin{equation}
    \mathcal{L}_{\mathrm{MSE}} =
        \frac{1}{|\Theta_{\mathrm{in}}|F}
        \sum_{f,\,\theta\in\Theta_{\mathrm{in}}}
        \!\bigl(D^{(f,\theta)} - T_{\mathrm{in}}\bigr)^{2}
\end{equation}
\begin{equation}
        \frac{1}{|\Theta_{\mathrm{out}}|F}
        \sum_{f,\,\theta\in\Theta_{\mathrm{out}}}
        \!\bigl(D^{(f,\theta)} - T_{\mathrm{out}}\bigr)^{2}.
\end{equation}

The MSE between the $D^{(f,\theta)}$, being the sound pressure level normalised to on-axis, and the target is evaluated for the in-coverage region and out-of-coverage region for the horizontal and vertical planes separately. 

\section{Results}\label{sec:results}

\subsection{Validation - Error}

\begin{figure}[h]
 \centerline{
 \includegraphics[width=7cm]{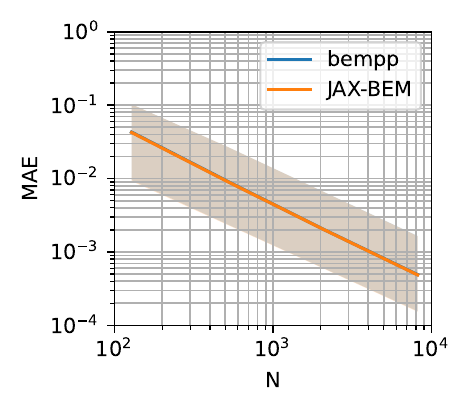}}
 \caption{Error vs. analytic solution for increasing mesh refinement. (Scattering from a rigid sphere)}
 \label{fig:error_n}
 \centerline{
 \includegraphics[width=7cm]{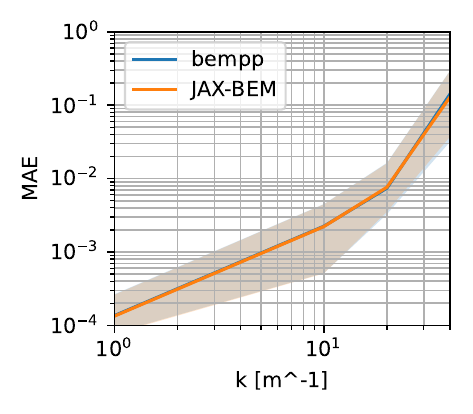}}
 \caption{Error vs. analytic solution for increasing wavenumber $k$. (Scattering from a rigid sphere)}
 \label{fig:error_k}
\end{figure}

The scattering from a rigid sphere in 3D was chosen as a validation problem, because it has analytic solutions in the form of the Mie series \cite{morse_ingard_1968}. The Burton-Miller formulation was used to avoid spurious interior resonances. \texttt{JAX-BEM} was validated by comparing the scattered field to \texttt{bempp} at a range of mesh sizes and wavenumbers. The mean absolute error of \texttt{JAX-BEM} relative to \texttt{bempp} ranged from $8\times10^{-4}$ ($N=128$) to $8.2\times10^{-6}$ ($N=8192$). Mean absolute error compared to analytic solutions ranged from $4\times10^{-2}$ ($N=128$) to $4\times10^{-4}$ ($N=8192$) for both \texttt{JAX-BEM} and \texttt{bempp}, demonstrating good agreement (Figs. \ref{fig:error_n}, \ref{fig:error_k}). 

Although \texttt{bempp} uses \texttt{numpy} which defaults to 64 bit precision (\texttt{complex128}) there was no significant difference in error to \texttt{JAX} using the default 32 bit precision (\texttt{complex64}), indicating that discretisation error dominates. 

\subsection{Wall clock time}

\begin{figure}[h]
 \centerline{
 \includegraphics[width=6cm]{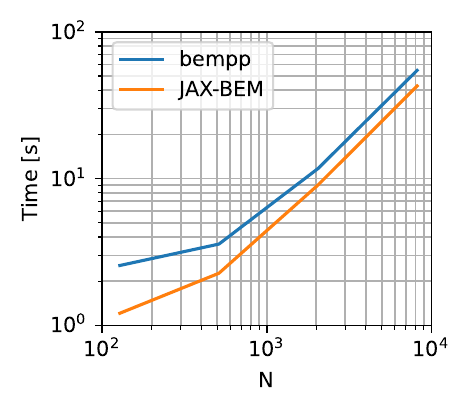}}
 \caption{Wall clock time for varying mesh refinement $N$ and $k=4$. \texttt{bempp} using \texttt{numba} JIT backend. \texttt{JAX} without calculation of gradients. }
 \label{fig:time_n}
\end{figure}

Fig. \ref{fig:time_n} shows the wall clock time vs number of elements $N$. There were $2\times10^{6}$ domain evaluation points on a $128^3$ 3D grid. The CPU was an AMD Ryzen™ AI Max+ 395. \texttt{bempp} is using the \texttt{numba} JIT backend, while \texttt{JAX-BEM} uses the built-in \texttt{XLA} backend. For both methods, timing was performed after a small initial warm-up solve to allow the JIT functions to compile. \texttt{JAX-BEM} likely performs slightly better due to its dense assembly, \texttt{bempp} uses a chunked approach. 

\subsection{Geometry optimisation}

We use reverse-mode automatic differentiation to optimise the directivity of a loudspeaker horn driven by constant unit velocity. A coupled formulation was used, consisting of a closed interior domain (horn) with rigid boundary conditions, radiating throat cap and fictitious mouth cap. The interior domain is terminated with an absorbing Dirichlet-to-Neumann map boundary condition at its mouth, data from which is fed into the unbounded exterior domain model. It is noted that other formulations may be more accurate.

\begin{figure*}[t]
 \centerline{
 \includegraphics[width=16.0cm]{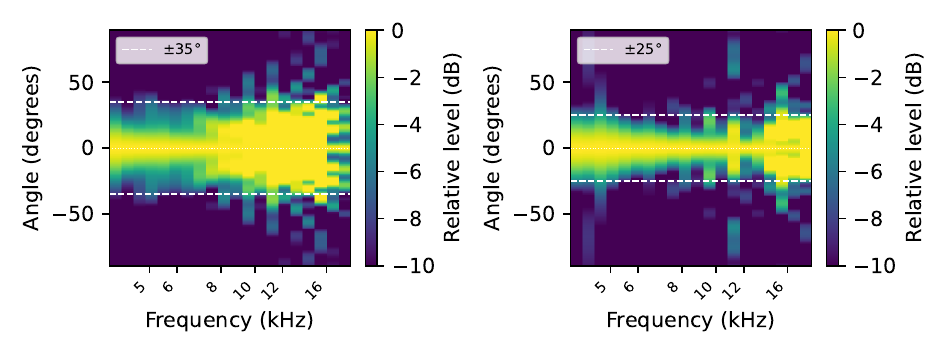}}
 \caption{Initial directivity (normalised to on-axis). Left: Horizontal, Right: Vertical.}
  \label{fig:directivity_init}
 \centerline{
 \includegraphics[width=16.0cm]{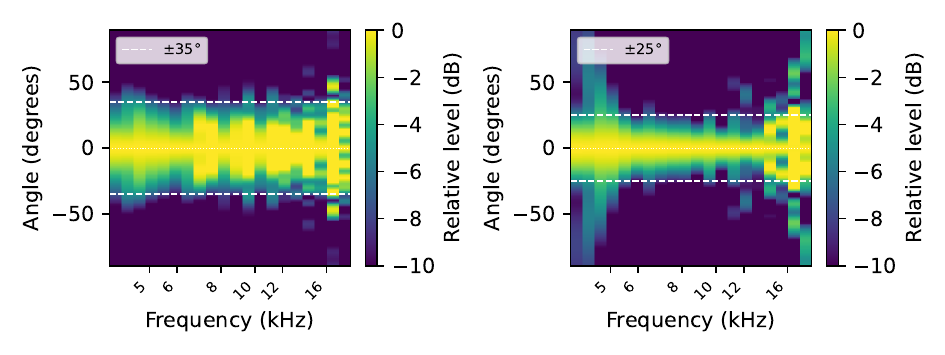}}
 \caption{Optimised directivity (normalised to on-axis). Left: Horizontal, Right: Vertical.}
 \label{fig:directivity_opt}
\end{figure*}

Loudspeaker horns are sensitive and difficult to design for wide bandwidth constant directivity. Therefore, gradient-based optimisation is particularly attractive for this application. Fig. \ref{fig:directivity_init} shows the normalised horizontal and vertical directivity maps of an unoptimised design (simple conical horn). An approximately constant directivity characteristic is apparent, but there is substantial diffraction above ~8 kHz in both planes. For the optimisation, the target directivity was set to $70\degree$ horizontal, $50\degree$ vertical. The solver ran in batches of 20 frequencies logarithmically spaced from 4 kHz to 18 kHz. The exterior domain was sampled using three arcs of observation points at 10 m spaced at $1\degree$ intervals across $90\degree$ in the horizontal, vertical and $45\degree$ planes. The small number of points in the domain meant this experiment was more efficient to perform on the CPU. 

\begin{figure}[h!]
 \centerline{
 \includegraphics[width=4.9cm]{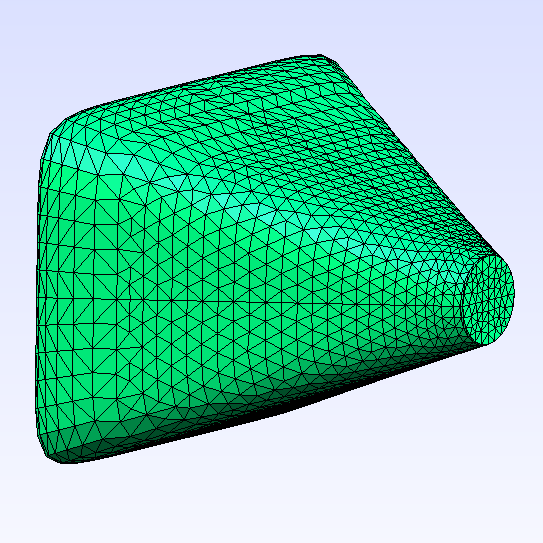}}
 \caption{Intitial horn mesh.}
 \label{fig:mesh_init}
  \centerline{
 \includegraphics[width=4.9cm]{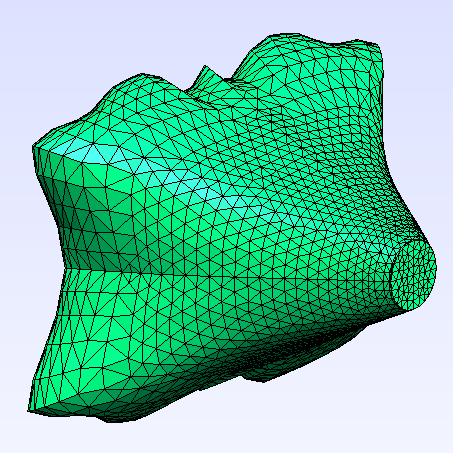}}
 \caption{Optimised horn mesh.}
 \label{fig:mesh_opt}
\end{figure}

Observing the differences between the initial mesh (Fig. \ref{fig:mesh_init}) and the optimised mesh (Fig. \ref{fig:mesh_opt}) several interesting features are apparent. The most significant is that the optimiser has arrived at a complex mouth shape. A smooth mouth termination has been added, and this varies from a very large flare at $45\degree$ to a sharper termination in the horizontal and vertical planes. A smooth edge termination contributes the most to the improvements seen in the directivity plot. 

Fig. \ref{fig:directivity_opt} shows the directivity plots of the optimised design, the diffraction has been reduced, and the general characteristic is that of more constant directivity in the horizontal plane. However, in the vertical plane, although the directivity has been widened to the target, the out of coverage sound pressure level has increased at some frequencies, indicating the difficulty of competing optimisation objectives. 

Although the mesh was small (a quadrant was used with symmetry planes, 948 elements) the memory usage was approximately 100 GB for 20 frequencies per iteration in 32 bit precision (\texttt{complex64}). The current implementation is by no means fully optimised, but this does highlight a fundamental difficulty with BEM in that it scales as $\mathcal{O}(N^2)$. H-Matrix compression \cite{Bruyninckx_2025} is a potential solution, if implemented in a differentiable framework. 

\section{Discussion and further work}\label{sec:discussion}

Validation of the forward algorithm and an optimisation example have been demonstrated. The most important area for further work is to rigorously validate the accuracy of gradients, as optimisers can still work with noisy or slightly inaccurate gradients. 

Multiple full BEM solves (several frequencies) per iteration is very expensive, and automatic gradient tracking increases memory usage. Therefore, a key avenue for further work is to explore methods by which a full re-solve can be avoided for each geometry update. Because each optimisation iteration is an incremental change to the mesh, it seems inefficient that the entire problem must be solved from scratch. Warm-starting \texttt{GMRES} with the previous solution provides a useful speed-up, but the forward algorithm remains expensive. 

There are several possible BEM formulations to model a loudspeaker horn. Gradient-based optimisation and implicit differentiation are new parameters that influence which choice of formulation is the most accurate and efficient. It was found that a Dirichlet-to-Neumann map formulation was more stable than a coupled subdomain formulation. There is extensive further work to be done in this area to understand the numerical behaviour in more detail, and determine the best formulation for both accuracy and gradient-based optimisation. 

Validation with acoustic measurements of 3D printed devices should also be performed. This should be combined with a comparison to conventional (non-gradient-based) optimisation in terms of performance and computation time. 

\section{Conclusions}\label{sec:conclusions}

We have demonstrated a fast, differentiable and GPU computable boundary element method implemented using the automatic differentiation framework \texttt{JAX}. The method is differentiable from the domain solution back to the mesh vertices, computing gradients of the loss function with respect to the input mesh geometry. This enables efficient optimisation of many input parameters using a scalar loss value with reverse-mode automatic differentiation. The method was validated using the problem of scattering from a rigid sphere and applied to the problem of optimising loudspeaker horn directivity. 

\section{Code}

The \texttt{JAX-BEM} code and validation example (scattering from a rigid sphere) can be downloaded from

\url{https://github.com/jrhip/jax-bem}

\section{Acknowledgments}
This work was supported by EPSRC and Funktion-One Research Ltd. as part of the CDT in Sustainable Sound Futures. 

\url{https://soundfutures.salford.ac.uk/}

\clearpage

\printbibliography
\end{document}